\documentclass[11pt,prd,superscriptaddress,nofootinbib]{revtex4}
\usepackage[usenames,dvipsnames]{xcolor}

\usepackage{amssymb,amsmath,amsfonts}

\begin{document}
\begin{titlepage}

\hfill\parbox{5cm} { }

\vspace{25mm}

\begin{center}
{\Large \bf  $ \rho$  meson-nucleon coupling constant from the soft-wall AdS/QCD model}

\vskip 1. cm
  {Narmin Huseynova $^{a,b}$\footnote{e-mail : nerminh236@gmail.com} and
  Shahin Mamedov$^{a,c}$\footnote{e-mail : sh.mamedov62@gmail.com}}

\vskip 0.5cm

{\it $^a\,$Institute for Physical Problems, Baku State University, Z.Khalilov 23, Baku, AZ-1148, Azerbaijan}\\
{\it $^b\,$Theoretical Physics Department, Physics Faculty, Baku State University, Z.Khalilov 23, Baku, AZ-1148, Azerbaijan}\\
{\it $^c\,$Department of Physics, Gazi University, Teknikokullar, 06100 Ankara, Turkey}\\
\end{center}

\thispagestyle{empty}

\vskip2cm


\centerline{\bf ABSTRACT} \vskip 4mm

\vspace{1cm}
 In the framework of the soft-wall model of AdS/QCD we calculated the $\rho$ meson  nucleon coupling constant. Bulk-to boundary propagators for the free vector and spinor fields are presented, whose boundary values correspond to $\rho$ meson and to the nucleon respectively. The interaction Lagrangian between these fields is written in the bulk of AdS space and includes magnetic type interactions as well. Using the AdS/CFT correspondence from the bulk interaction action we derived the $g_{\rho NN}$ coupling constant in the boundary QCD. We found that the soft-wall model result for the $g_{\rho NN}$ constant is  more close to empirical values than the one obtained in the hard-wall model.
\vspace{2cm}


\end{titlepage}

\section{introduction}

The AdS/CFT correspondence~\cite{1,2,3} discovered in recent decades is successfully applied for solving problems in different branches of physics. This correspondence establishes duality between the fields in the bulk of an anti-de~Sitter (AdS) space with the field theory operators defined on the ultraviolet (UV) boundary of that space.
For Quantum Chromodynamics this duality has special importance, since the ordinary perturbation theory does not work at a low energy limit and it needs non-perturbative methods for solving problems of strong interactions in this energy region. Application of AdS/CFT correspondence principle (or holography principle) to the QCD theory, which is named as holographic QCD, found to be a useful concept for the studies in QCD at low energies.
In holographic QCD there are two approaches, which were named top-down and bottom-up ones.
 The top-down approach includes QCD models based on the string and $D$-brane theories. Meanwhile, the bottom-up approach of the holographic QCD is based on direct application of the AdS/CFT principle to the theory of strongly interacting particles and so is frequently called  AdS/QCD. Two main models of AdS/QCD, which are known as the hard-wall and soft-wall models~\cite{4,5,6,7,8,9,10,11}, were constructed under the finiteness condition of the action for the model at the infrared (IR) boundary of AdS space. In the hard-wall model this condition is ensured by the cut off at this boundary and in the soft-wall one the condition is satisfied by introducing an exponential factor, which suppresses expressions under the integral over the extra dimension at infinite values of this dimension ~\cite{12,13,14}. The hard-wall model produces a linearly growing mass spectrum of mesons and the soft-wall one gives linearly growing spectrum for the squared mass. Both models are a powerful tool for the calculation of various coupling constants of the strongly interacting particles as well as the form factors for them. In the AdS/QCD models framework these coupling constants can be derived for both the ground and excited states of mesons and baryons~\cite{10,15,16,17}, which enlarges the range of applicability of these models. There is light-front approach in AdS/QCD, within which these coupling constants and form factors could be derived as well ~\cite{18}. Besides calculations performed in vacuum, the AdS/QCD models are applied to research of physical effects and quantities in the dense nuclear medium~\cite{19,20,21,22}.

 Studies of mesons in the AdS/QCD framework have been started from the pioneering work~\cite{5} and the $\rho$ mesons, as lightest ones in the vector meson series, were considered in several papers~\cite {9,23,24,25}. Here we aim to consider the $\rho $ meson strong coupling constant with nucleons in the framework of the soft-wall model in vacuum. Note, that this coupling constant has been evaluated in the framework of the hard-wall model in~\cite{10,16,25} and in top-down approach in~\cite{26,27,28}. For this coupling constant there is empirical data as well, that give rise to a possibility to make  comparison with it. Such an analysis may serve as an additional verification of the soft-wall model on the example of computing of the $g_{\rho NN}$
coupling constant.

The present paper was arranged as follows: First, we define the bulk geometry of the model, then we briefly present the equations of motion for the free vector, scalar and spinor fields in the bulk and write down bulk-to-boundary propagators for these fields. Next, we write a Lagrangian for the vector-spinor interaction in this geometry and derive the boundary $g_{\rho NN}$ coupling constant from the bulk Lagrangian. Finally, we give a numerical results for the coupling constant and carry out comparison of the obtained values with the empirical one.

\section{The soft-wall model}

The action for the soft-wall model in general can be written in the form:
\begin{equation}
S=\int d^4xdz\sqrt{g}e^{-\Phi(z)}\mathcal{L}\left(x,z\right),
\label{1}
\end{equation}
where $g=|\det g_{MN}|$  $(M,N=0,1,2,3,5)$ and the extra dimension $z$ varies in the range $\epsilon\leq z<\infty$ $\left(\epsilon\rightarrow0\right)$.
 An exponential factor was introduced to make the integral over the $z$ finite at IR boundary $(z\rightarrow \infty)$
and the dilaton field  $\Phi\left(z\right)=k^2 z^2$ is chosen in the usual form for the soft-wall model.
The parameter $k$ is a free one and it is usually fixed by  fitting mass spectrum with the experimental data for the particles considered in the model. The metric of AdS space is given in Poincare coordinates and its radius $R$ has been set equal to 1:
\begin{equation}
ds^2=\frac{1}{z^2}\left(-dz^2+\eta_{\mu\nu}dx^{\mu}μdx^{\nu}\right),\quad \mu,\nu=0,1,2,3.
\label{2}
\end{equation}
The 4-dimensional metric $\eta_{\mu\nu}$ has Minkowski signature:
\begin{equation}
\eta_{\mu\nu}=diag(1,-1,-1,-1).
\label{3}
\end{equation}

\subsection{$\rho$-mesons in soft-wall model}

Let us introduce in the bulk of AdS space two gauge fields $A_L^M$ and $A_{R}^{M}$, which transform as a left and right chiral fields under $SU(2)_L\times SU(2)_R$ chiral symmetry group of the model. The chiral symmetry group is broken to the isospin group $SU(2)_V$ (in vector representation) due to the interaction of the bulk gauge fields with
the scalar field $X$. According to the AdS/CFT correspondence the bulk $SU(2)_V$ symmetry group is a symmetry group of the dual boundary theory, {\it i.e.} isospin symmetry of QCD and the $\rho$ meson triplet is described by this representation of the $SU(2)_V$ group. From the gauge fields $A_L^M$ and $A_{R}^{M}$ one can construct a bulk vector
$V^M=\frac{1}{\sqrt{2}}\left(A_L^M +A_{R}^{M}\right)$ and the axial vector  $A^M=\frac{1}{\sqrt{2}}\left(A_L^M -A_{R}^{M}\right)$ fields. According to the AdS/CFT correspondence of vector field the UV boundary value of the Kaluza-Klein modes of the bulk vector field corresponds to the vector meson series of the dual theory.
Since the $\rho$ meson is the lightest vector meson in particle physics it corresponds to the first state of these modes. The action for the gauge field sector will be written in terms of bulk vector and axial-vector fields as follows:
\begin{equation}
  S_{gauge}=-\frac{1}{4g_5^2}\int d^5x\sqrt{g}e^{-\Phi(z)}Tr\left[F_L^2+F_R^2\right]
   =-\frac{1}{4g_5^2}\int d^5x\sqrt{g}e^{-\Phi(z)}Tr\left[F_V^2+F_A^2\right],
\label{4}
\end{equation}
where the field stress tensor is $F_{MN}=\partial_MV_N-\partial_NV_M-i\left[V_M,V_N\right]$, $V_M=V_M^at^a$,
$t^a=\sigma^a/2$ and $ \sigma^a$ are Pauli matrices. 5-dimensional coupling constant $g_5$ is related to the number of colors $N_c$ in the dual theory as $g_5^2=12\pi^2/N_c$ ~\cite{5}. We choose the axial-like gauge, $V_5=0$, to fix the $V_5$ component~\cite{5}.
 4-dimensional components of the vector field at UV boundary $\left( V_{\mu}(x,z=0)\right)$ correspond to the source of the vector current. Fluctuations of the bulk vector field correspond to the vector mesons at the boundary. Since the boundary value of the vector field $V_M$ corresponds to the vector mesons, only the transverse part of this bulk vector field $\left(\partial^{\mu}V_{\mu}^T=0\right)$ is interesting for our study. It is useful to write the transverse part of the bulk vector field $V_{\mu}^T(x,z)$  in momentum space by help of Fourier transformation\footnote{Hereafter we omit the sign of transversity $T$ on $V_M^T.$}. The equation of motion for Fourier components $\widetilde{V}_{\mu}^a(p,z)$ is easily obtained from the action (\ref{4}) and has the form
\begin{equation}
\partial_z\left[\frac{1}{z}e^{-k^2z^2} \partial_z\widetilde{V}_{\mu}^a(p,z)\right]+p^2\frac{1}{z} e^{-k^2z^2}\widetilde{V}_{\mu}^a(p,z)=0.
\label{5}
\end{equation}
The $\widetilde{V}_{\mu}^a(p,z)$ can be written as $\widetilde{V}_{\mu}^a(p,z)=V_{\mu}^a(p)V(p,z) $ and at UV boundary
$V(p,z) $ satisfies the condition $V(p,\epsilon)=1$, since it is assumed that 4-dimensional physical world resides at this boundary of the AdS space. For the $n$-th mode $V_n(z)$ in the Kaluza-Klein decomposition $V(p,z)=\sum\limits_{n=0}^{\infty}V_n(z)f_n(p) $ with mass $m_n^2=p^2$ the equation (\ref{5}) obtains the form:
\begin{equation}
\partial_z\left(e^{-B(z)} \partial_z V_n \right)+m_n^2 e^{-B(z)} V_n=0,
\label{6}
\end{equation}
where $B (z)=\Phi(z)-A(z)= k^2z^2+\ln z$. Making substitution
\begin{equation}
 V_n(z)=e^{B(z)/2}\psi_n(z)
\label{7}
\end{equation}
 the equation (\ref{6}) is reduced to the Schroedinger equation form and has a solution~\cite{7} in terms of Laguerre polynomials $L_m^n$:
\begin{equation}
\psi_n(z)=e^{-k^2 z^2/2}\left(kz\right)^{m+1/2}\sqrt{\frac{2n!}{\left(m+n\right)!}}L_n^m\left(k^2z^2 \right).
\label{8}
\end{equation}
For the eigenvalues $m_n^2$ there is a linear dependence on the number $n$:  $m_n^2=4k^2(n+1)$, which enables us to fix the free parameter $k$. In the AdS/CFT correspondence $m_n^2$ is identified with the mass spectrum of the vector mesons
in the dual boundary QCD. For the $\rho$-meson we have $m=1$ and the $V_n(z)$ becomes~\cite{8,9}
\begin{equation}
V_n(z)=k^2z^2\sqrt{\frac{2}{n+1}}L_n^1\left(k^2z^2 \right).
\label{9}
\end{equation}

 It should note that solution (\ref{9}) was obtained for the free field case and does not take into account the back reaction of the bulk spinor field, which we will introduce in the next paragraph in order to describe nucleons on the boundary.

\subsection{Axial vector and pseudoscalar fields and chiral symmetry breaking}

Let us now consider axial vector field $A^M$. This field is related to pseudoscalar field $X$, which is introduced into the AdS/QCD models in order to perform breaking of the chiral $SU(2)_L\times SU(2)_R$ symmetry by Higgs mechanism~\cite{7}. The $X$ field transforms under bi-fundamental representation of this symmetry group and action
for this field has the form:
\begin{equation}
  S_{X}=\int d^5x\sqrt{g}e^{-\Phi(z)}Tr \{|DX|^2+3|X|^2 \},
\label{10}
\end{equation}
where $-3$ is the squared $5$-dimensional mass of the X field ($M_5^2=-3$). Covariant derivative $D^M $ is defined as $D^M X=\partial^M X-iA_L^M X+XA_R^M$ and contains  interaction with the axial vector field. So, the equation of motion for the axial vector Kaluza-Klein modes $A_n(z)$ takes into account interaction with the $X$ field~\cite{5}:
\begin{equation}
\partial_z\left(e^{-B(z)}\partial_zA_n\right)+\left[ m_n^2-g_5^2e^{2A(z)}X(z)^2\right]e^{-B(z)}A_n=0
\label{11}
\end{equation}
and the equation of motion for the $X$ field was obtained as below:
\begin{equation}
\partial_z\left(e^{-B(z)+2A(z)}\partial_z X(z)\right)+3e^{-B(z)+4A(z)}X(z)=0.
\label{12}
\end{equation}
 It is noteworthy that higher order interaction terms may be included into the action for the $X$ field for accuracy, however we shall only consider the (\ref{10}). Also, for our calculation of $g_{\rho NN}$ we shall be content with the
following asymptotic solution\footnote{More detailed analysis of the asymptotic solutions for the $X(z)$ can be found in~\cite{7}.}
of (\ref{12}) at $z\rightarrow 0$ limit:
\begin{equation}
X(z)\approx\frac{1}{2}m_qz+\frac{1}{2}\sigma z^3=v(z).
\label{13}
\end{equation}
Here the coefficient $m_q$ is the mass of $u$ and $d$ quarks and $\sigma$ is the value of chiral condensate. The coefficients $m_q$ and $\sigma$ were established from the UV and IR boundary conditions imposed on the solution for the field $X$. At $z\rightarrow \infty$ this solution is suppressed by an exponential factor in the action.

\subsection{Nucleons in soft-wall model}

Nucleons were introduced into the soft-wall model in~\cite{8} and their excited states within this model were considered in~\cite{12,15}. In the hard-wall model nucleons were introduced in~\cite{11} on the basis of the AdS/CFT correspondence for the spinor field~\cite{29}. Following~\cite{8} and~\cite{15} we shall present here some formulas of profile function derivation for the spinor field in the soft-wall model. It should be noted that a nucleon doublet in the hard-wall model is described by a pair of bulk fermions, because the left- and right-handed components of the nucleon operator at the boundary are described by two bulk fermion fields having opposite signs of 5-dimensional mass $M$ (detailed substantiation may be found in~\cite{11}. However, the soft-wall model  Lagrangian contains additional term $\Phi\overline{\Psi}\Psi $, which describe coupling of a dilaton field with the bulk fermion fields ~(\cite{8}) and the sign of this term for the second fermion field is chosen oppositely to the one for the first fermion \cite{11}. So, in order to describe the nucleon doublet in the boundary QCD
it is necessary to introduce two bulk fermions $N_1$ and $N_2$ having opposite signs of $M$ and  $\Phi\overline{\Psi}\Psi $ term, then eliminate extra chiral components at the UV boundary by the boundary conditions. Let us demonstrate profile function derivation for the fermion field $\Psi_1(x,z)$ in the bulk of the AdS space (\ref{2}). The action for this field without taking into account interaction with the gauge fields, may be written as follows:
\begin{equation}
S_{F_1}=\int d^4xdz\sqrt{g}e^{-\Phi(z)}\left[ \frac{i}{2}\overline{\Psi}_1e_A^N\Gamma^AD_N\overline{\Psi}_1
  -\frac{i}{2}\left(D_N\Psi_1\right)^\dagger \Gamma^0e_A^N\Gamma^A\Psi_1-(M+\Phi(z))\overline{\Psi}_1\Psi_1 \right],
\label{14}
\end{equation}
where $e_A^N=z\delta_A^N$ is the inverse vielbein and the covariant derivative is
$D_N=\partial_N+\frac{1}{8}\omega_{N AB}\left[\Gamma^A,\Gamma^B\right]$.
Non-zero components of spin connection are: $\omega_{\mu z\nu }=-\omega_{\mu\nu z}=\frac{1}{z}\eta_{\mu\nu}$.
The 5-dimensional matrices $\Gamma^A$ are defined as $\Gamma^A=(\gamma^\mu, -i\gamma^5)$ and obey the anticommutation relation $ \{\Gamma^A,\Gamma^B\}=2\eta^{AB}$. Note that the mass term in this action contains an additional term
with $\Phi(z)$ that accounts for the dilaton-fermion interaction, while the back reaction of other fields existing in the bulk of the AdS space were ignored in (\ref{14}) and we shall use in our calculations the profile function of fermion in this approximation. The equation of motion obtained from the action (\ref{14}) has the form:
\begin{equation}
 \left[ ie_A^N\Gamma^AD_N-\frac{i}{2}(\partial_N\Phi)e_A^N\Gamma^A-(M+\Phi(z))\right]\Psi_1=0.
\label{15}
\end{equation}
It is convenient to solve (\ref{15}) for $\Psi$ in terms of left- and right-handed components defined as $\Psi_{L,R}=\left(1/2\right)\left(1\mp\gamma^5\right)\Psi$. In  momentum space $\Psi_{L,R}$ are written as product of the UV boundary fields $\Psi_{L,R}^0(p)$, profile functions $f_{L,R}(p,z)$ and the factor of conformal dimension $\Delta$, {\it i.e.} as $\Psi_{L,R}(p,z)=z^{\Delta}\Psi_{L,R}^0(p)f_{L,R}(p,z)$. The boundary fields $\Psi_{L,R}^0(p)$ are related to each other by the 4-dimensional Dirac equation $\not{\!}p\Psi^0_R(p)=p\Psi^0_L(p)$, since spinors in 5 dimension are the ones in 4 dimension. The equation ~(\ref{15}) will give us relations between profile functions $f_{L}(p,z)$ and $f_{R}(p,z)$ \cite{8}:
\begin{eqnarray}
\left(\partial_z-\frac{d/2-\Delta+\Phi+\left(M+\Phi\right)}{z}\right)f_{1R}=-pf_{1L}, \nonumber \\
\left(\partial_z-\frac{d/2-\Delta+\Phi-\left(M+\Phi\right)}{z}\right)f_{1L}=pf_{1R}.
\label{16}
\end{eqnarray}
Using the relation $\Delta=\frac{d}{2}-M$ known form the AdS/CFT correspondence (with $d=4$ for our model), it is obtained from (\ref{16}) the second order differential equations for the profile functions $f_{L,R}(p,z)$
\begin{eqnarray}
\left[\partial_z^2-\frac{2}{z}(M+k^2z^2) \partial_z+\frac{2}{z^2}(M-k^2z^2)+p^2\right]f_{1R}=0, \nonumber \\
\left[\partial_z^2-\frac{2}{z}(M+k^2z^2)\partial_z+p^2\right]f_{1L}=0.
\label{17}
\end{eqnarray}
The $n$-th normalized Kaluza-Klein mode $f^{(n)}_{L,R}\left( z\right)$ of the solutions $f_{L,R}$ with $p^2=m_n^2$ may be expressed in terms of Laguerre polynomials $L_{n}^{(\alpha)}$:
\begin{eqnarray}
f_{1L}^{(n)}(z)&=&n_{1L}\left(kz\right)^{2\alpha}L_{n}^{(\alpha)}\left(kz\right),\nonumber \\
f_{1R}^{(n)}(z)&=&n_{1R}\left(kz\right)^{2\alpha-1}L_{n}^{(\alpha-1)}\left(kz\right).
\label{18}
\end{eqnarray}
Parameter $\alpha$ is related to the 5-dimensional mass $M$ via $\alpha=M+\frac{1}{2}$.
It relates the mass of the $n$-th mode $m_n$ to the number $n$ in the following $m_n^2=4k^2\left(n+\alpha\right)$, which serves as another condition to fix parameter $k$ of the model. The constants $n_{L,R}$ are found from the normalization condition
\begin{equation}
\int dz \frac{e^{-k^2z^2}}{z^{2M}}f_{1L}^{(n)}f_{1L}^{(m)}=\delta_{nm} \nonumber
\end{equation}
and are equal to
\begin{eqnarray}
n_{1L}&=&\frac{1}{k^{\alpha-1}}\sqrt{\frac{2\Gamma(n+1)}{\Gamma(\alpha+n+1)}},\nonumber \\
n_{1R}&=&n_{1L}\sqrt{\alpha+n}.
\label{19}
\end{eqnarray}
$M$ is equal to $M=\frac{3}{2}$ and hence $\alpha=2$.

For the second bulk fermion field we have to change the sign of $\left(M+\Phi(z)\right)$ term in the Lagrangian to opposite, ({\it i.e.} make a replacement $\left(M+\Phi(z)\right)\rightarrow -\left(M+\Phi(z)\right)$ in (\ref{14})),
because, as was noted above, this fermion is needed for the description of another chiral component of the nucleons at the UV boundary. There are following relations between the profile functions of the first and second bulk fermion fields~\cite{11,15}:
\begin{equation}
f_{1L}=f_{2R},\quad f_{1R}=-f_{2L}\label{20},
\end{equation}
which could be deduced from (\ref{16}) as well performing the above replacement and $1\rightarrow 2$ in it\footnote{The relation (\ref{20}) is not unique~(\cite{11}).}.

\section{Bulk interaction and the $g_{\rho NN}$ coupling constant}

 A boundary meson-nucleon coupling constant will be derived from the 5-dimensional action for the interaction in the bulk of the AdS space between the fermion fields $N_1$ and $N_2$ with vector field $V_{\mu}$:
\begin{equation}
 S_{int}=\int d^4x dz e^{-\Phi(z)}\sqrt{g} \mathcal{L}_{int}.
\label{21}
\end{equation}
According to the AdS/CFT correspondence classical bulk action $S$ is the generating function $Z$ for the vacuum expectation value of the current in dual 4-dimensional theory at the UV boundary. For our problem this principle will be written as
\begin{equation}
<J_{\mu}>=-i\frac{\delta Z_{QCD}}{\delta \tilde{ V}_{\mu}^0}|_{\tilde{ V}_{\mu}^0=0}
\label{22}
\end{equation}
with
\begin{equation}
Z_{QCD}=e^{iS_{int}},
\label{23}
\end{equation}
where $\tilde{V}_{\mu}^0=\tilde{V}_{\mu}(q, z=0)=V_{\mu}(q)$ is the UV boundary value of the vector field $\left(V(z=0)=1\right)$ and $J_{\mu}$ obtained from the variation of $exp\left(iS_{int}\right)$ will be identified with the nucleon current\footnote{We did not take into account the isospin structure in the current, since in the vacuum case all $g_{\rho NN}$ constants are the same and this does not matter. However, in the case of medium having non-zero average value of isospin the isospin structure has importance~\cite{22}.} $J_{\mu}(p^{\prime},p)=g_{\rho NN}\bar{u}(p^{\prime})\gamma_{\mu}u(p)$
in the QCD theory, which exists at this boundary of the AdS space.
$\tilde{V}_{\mu}^0$ is the source for the $J_{\mu}$ current.
There is an energy-momentum conservation relation between 4-momenta $q$, $p^{\prime}$ and $p$: $q=p^{\prime}-p$,  which arises as a result of integration over space-time coordinates $x$. $p^{\prime}$ and $p$ on the AdS gravity side are four momenta of the bulk spinor fields after and before the interaction with the vector field,  respectively. But on the QCD side these are four momenta of the final and initial nucleon respectively.
The integral over $z$ arising on the right-hand side of (\ref{22}) will be accepted as the $g_{\rho NN}$ coupling constant existing in the nucleon current $J_{\mu}(p^{\prime},p)$
when two currents, the fermion current at the boundary and the nucleon current, are identified as above.

Generally, the interaction Lagrangian $\mathcal{L}_{int} $ is constructed basing on the gauge invariance of the model and it contains different kinds of interaction terms~\cite{16,10}. On constructing the interaction Lagrangian for deriving the
$g_{\rho NN}$ coupling constant we may follow \cite{16}, where this was done within the hard-wall model framework. Since the Lagrangian terms in hard-wall model were constructed on a gauge invariance requirement and they are not model dependent, it is correct to apply these terms in the soft-wall model case as well. First, $\mathcal{L}_{int}$ contains a term of minimal gauge interaction of the vector field with the current of fermions
\begin{equation}
 \mathcal{L}_{\rho NN}^{(0)}=\overline{N}_{1} e_{A}^M\Gamma^{A}V_{M}N_{1}+\overline{N}_{2}e_{A}^M\Gamma^{A}V_{M}N_{2},
\label{24}
\end{equation}
where $\Gamma^{A}$ matrices are the ones in flat tangent space and were defined in the previous section. After performing the integrals in momentum space and applying the holography principle this Lagrangian term gives the following contribution to the $g_{\rho NN}$ constant represented in terms of integral over $z$:
\begin{equation}
  g_{\rho NN}^{(0)nm}=\int_{0}^{\infty}\frac{dz}{z^{4}}e^{-\Phi(z)}V_{0}(z)
    \left( f_{1L}^{(n)*}(z)f_{1L}^{(m)}(z) + f_{2L}^{(n)*}(z)f_{2L}^{(m)}(z)\right),
\label{25}
\end{equation}
where the relations (\ref{20}) has been used for shortness. In (\ref{25}) $ V_{0}(z)=k^2z^2\sqrt{2}L_0^1\left(k^2z^2 \right)$ is the profile function for zero Kaluza-Klein mode of the vector field, the UV boundary value of which corresponds in the AdS/CFT correspondence to the ground-state of $\rho$ meson and the superscript indices $n$ and $m$  indicate the number of excited states of the initial and final nucleons respectively.

The bulk spinor fields have magnetic moments and they may interact with the bulk vector field by means of these moments. It is obvious this kind of interaction will contribute to the $g_{\rho NN}$ coupling constant in the boundary QCD as well. The Lagrangian for this interaction $L_{FNN}^{(1)}$
\begin{equation}
 L_{FNN}^{(1)}=ik_{1}e_{A}^{M} e_{B}^{N} \left[ \overline{N}_{1}\Gamma^{AB}(F_{L} )_{MN}N_{1}-\overline{N}_{2}\Gamma^{AB}(F_{R})_{MN}N_{2}\right]
\label{26}
\end{equation}
 was constructed in ~\cite{16} and its contribution to $g_{\rho NN}$ was calculated within the hard-wall model. Besides this simple fermion-vector field interaction by means of magnetic moment there is another, but more complicated interaction of these fields by the magnetic moment where the bulk scalar $X$  takes part as well. The Lagrangian for this interaction which also was written in ~\cite{16} has the form:
\begin{equation}
 L_{FNN}^{(2)}=\frac{i}{2}k_{2}e_{A}^{M}e_{B}^{N}\left[ \overline{N}_{1}X\Gamma^{AB}(F_{R})_{MN}N_{2}+\overline{N}_{2}X^{+}\Gamma^{AB}(F_{L})_{MN}N_{1}-h.c.\right].
\label{27}
\end{equation}
Since the $X$ field was included into the interaction this term was constructed as one changing chirality of the boundary nucleons and the gauge field  in the bulk interact with the fermions through the magnetic moment of last. So, on the boundary QCD side this term describes the $\rho$ meson coupling with the nucleon current through the magnetic moment of nucleons. We shall see later that in spite of the $X$ field is present in the Lagrangian (\ref{27}) there is no pion fields in the final expression of the action for the Lagrangian (\ref{27}). More precisely, the action for this Lagrangian contains non-zero term which is a zeroth order on the pion field term and so, it was not included into the $g_{\rho \pi NN}$ coupling constant but the $g_{\rho NN}$ one. Inclusion of terms (\ref{26}) and (\ref{27}) into the interaction Lagrangian $ \mathcal{L}_{int}$ can give us information about that how much contribute the magnetic moment interactions into the $g_{\rho NN}$ coupling constant. The $k_{1}$ and $k_{2}$ constants were found in~\cite{16} from fitting the $g_{\rho NN}$ and  $g_{\pi NN}$ coupling constants with their experimental values in the ground-state of nucleons. In spite of $L_{FNN}^{(1)}$ and $ L_{FNN}^{(2)}$ include both the vector and axial-vector fields we shall use their vector field part. The $\Gamma^{MN}F_{MN}$ matrix in (\ref{26}) and in (\ref{27}) contains two kind of terms which are $\Gamma^{5\nu}F_{5\nu}$ and  $\Gamma^{\mu\nu}F_{\mu\nu}$. It is useful to present the contributions made by these terms separately. Action integral (\ref{21}) of the first kind terms in the total Lagrangian $L_{FNN}=L_{FNN}^{(1)}+L_{FNN}^{(2)}$ is expressed in terms of mode profile function derivative $V'_{0}(z)$ and gives a final result for the $g_{\rho NN}$ constant in the following expression:
\begin{equation}
  g_{\rho NN}^{(1)nm}=-2\int_{0}^{\infty}\frac{dz}{z^{3}}e^{-k^2z^2}V'_{0}\left(z\right)\left[ k_{1}
  \left( f_{1L}^{(n)*}f_{1L}^{(m)}-f_{2L}^{(n)*}f_{2L}^{(m)}\right)
  +k_{2}v(z)\left( f_{1L}^{(n)*}f_{2L}^{(m)}+f_{2L}^{(n)*}f_{1L}^{(m)}\right)\right].
\label{28}
\end{equation}
The second kind terms, which are magnetic moment type, lead to the next contribution to the $g_{\rho NN}$:
\begin{equation}
  f_{\rho}^{nm}=4 m_N\int_{0}^{\infty}\frac{dz}{z^{3}}e^{-k^2z^2}V_{0}\left(z\right)\left[ k_{1}\left( f_{1L}^{(n)*}f_{1R}^{(m)}-f_{2L}^{(n)*}f_{2R}^{(m)}\right)
  +k_{2}v(z)\left( f_{1L}^{(n)*}f_{2R}^{(m)}+f_{2L}^{(n)*}f_{1R}^{(m)}\right)\right].
\label{29}
\end{equation}
These integrals have the same form as the ones obtained in the hard-wall model, except for the exponential factor and different expressions for the profile functions. So, following \cite{16}, we shall keep the description of the $ g_{\rho NN}$ constant as the sum of two terms, first of which  is $ g_{\rho NN}^{s.w.} =g_{\rho NN}^{(0)nm} +g_{\rho NN}^{(1)nm}$ and correspond to the coupling due to the "charge" and the second one, $f_{\rho}^{nm}$, can be considered as the contribution made by interaction of the $\rho$ meson with the nucleons via its magnetic moment.
It is recommended to carry out numerical analysis of these terms separately.

\section{Numerical analysis}

In order to get the value of the $g_{\rho NN}$ coupling constant within the model considered here we have to calculate the integrals for the expressions of $ g_{\rho NN}^{(0)nm} $, $ g_{\rho NN}^{(1)nm}$ and $f_{\rho}^{nm}$, which were presented in equations (\ref{25}), (\ref{28}) and (\ref{29}) respectively.
We calculate these integrals by use of MATHEMATICA package and then compare obtained results of the values of $ g_{\rho NN}^{nm}$ with the experimental data for this coupling constant and with the ones obtained within the other models.
For this aim free parameters $k$, $k_1$, $k_2$, $m_q$ and $\sigma$ existing in this model should be fixed. For the $k$ parameter it was found two values $k=0.350$~GeV in \cite{8} and $k=0.389$~GeV~\cite{30}.
These values were determined from matching of the mass spectra of the $\rho$ meson and nucleons which were derived in the soft-wall model with experimental values of the masses of these particles. Parameters $k_1$ and $k_2$ were fixed in \cite{16} calculating the $g_{\rho NN} $ and $g_{\pi NN}$ coupling constants in the hard-wall model framework. Constant $g_{\pi NN}$ can not be evaluated within the soft-wall model because of  the absence of explicit expression for the profile function for the pions in this model. However, since interactions (\ref{26}) and (\ref{27}) are model independent we may apply these values for the $k_1$ and $k_2$ constants in our calculations as well.

Though there is no data of the direct measurements of the $g_{\rho NN} $ coupling constant in the literature, this constant could be determined from other measurements in the experiment. For instance, in~\cite{31} the value of this constant $g_{\rho NN}^{exp}=2,52\pm0.06$ was estimated from the experimental data for width $\Gamma$ of the $\rho^0 \rightarrow e^+ e^-$ decay in ~\cite{32}. In~\cite{10,27,28} a range of $g_{\rho NN}^{emp} =4.2\sim6.5$ was used for this constant, which was empirically established in the~\cite{35}.  The hard-wall AdS/QCD model results are the following ones: $g_{\rho NN}^{h.w.} =-4.3\sim -6.2$ in ~\cite{10} and $g_{\rho NN}^{h.w.} =-8.6$ in~\cite{16}. The values obtained in the framework of QCD models are the next ones: value of $g_{\rho NN}^{q.s.r.} =-2.5\pm 1.1$ was obtained in~\cite{33} in the result of  application of QCD sum rules, while the value $g_{\rho NN}^{c.q.m.} =2.8$ was obtained in the chiral quark model framework in~\cite{34} and used in~\cite{16} for comparison with the hard-wall results.

Two different sets of values for $\sigma$ and $m_q$ were applied in studies carried out in the framework of AdS/QCD models in~\cite{31}: 1) the values of $\sigma=(0.368)^{3}$   GeV$^3$ and  $ m_q=0.00145$  GeV were found from fitting of the $\pi$ meson mass $m_{\pi}$ and its decay constant $f_{\pi}$ obtained in the soft-wall model framework with their experimental values and 2) the values of $\sigma=(0.327)^{3}$  GeV$^3$ and  $m_q=0.0023$  GeV were found in this work from fitting the results for these constants obtained in the hard-wall model with their experimental values.

In order to keep the clarity about relative contributions of different terms of Lagrangian, in Tables \ref{tab:1}--\ref{tab:4} we present the results for the $ g_{\rho NN}^{(0)nm} $, $ g_{\rho NN}^{(1)nm}$ and $g_{\rho NN}^{s.w.}$ coupling constants separately.
The numerical results for the set 1 are presented in Tables \ref{tab:1}--\ref{tab:2},
and for Set 2 are shown in Tables \ref{tab:3}--\ref{tab:4}.
Comparison of the soft-wall model results with the empirical values provided above and with the values obtained within the hard-wall model shows that for all values of parameters the soft-wall model gives the results more close to the experimental data than the hard-wall model. We can also notice that the contribution of $g_{\rho NN}^{(1)}$ to $g_{\rho NN}^{s.w.} $ is three times larger than the one coming from $g_{\rho NN}^{(0)}$, {\it i.e.} the interaction Lagrangians (\ref{26}) and (\ref{27}) make the main contribution.

In the ground-state of nucleons the tensorial coupling constant $f_{\rho}$ in the soft-wall model gets a value close to the one in the hard-wall model ~\cite{16}, but both values are larger than the value quoted in \cite{34}. However, for the excited states of nucleons the difference between the results of the hard-wall and the soft-wall models increases as the excitation number $n$ increases. Unfortunately, there is no experimental data for this coupling constant for $n\neq0$ and so, we can't make a conclusion about which model predicts a reliable result. From the comparison of tables it can also be seen that the $g_{\rho NN}$ constant is more sensitive to the value of parameter $k$ than to $\sigma$ and $m_q$.

\begin{table}[!htb]
\begin{center}
\begin{tabular}{|c|c|c|c|c|c|c|c|c|c|c|c|}
\hline
n & $m_N$  (GeV) & $m_N^{s.w.}$  (GeV) &$g_{\rho NN}^{(0)}$  & $g_{\rho NN}^{(1)}$ & $g_{\rho NN}^{s.w.} $ & $g_{\rho NN}^{exp} $ & $g_{\rho NN}^{h.w.} $&$g_{\rho NN}^{q.s.r} $  & $g_{\rho NN}^{c.q.m.} $ &$f_{\rho}^{s.w.}$ & $f_{\rho}^{h.w.}$  \\
\hline
0 & 0.94 & 1.089 & 1.64 & 5.14 & 6.78 & 2.52$\pm$0.06 & -8.6&-2.5$\pm$1.1&2.8 & -24.64 & -21.10 \\
 & & & & & & 4.2$\sim$6.5 & -4.3$\sim$-6.2 & & &&\\
\hline
1 & 1.44 &1.323& 2.14 & 11.95 & 14.09 & --- & 25.94&-&- & -107.33 & -20.18 \\
\hline
2 & 1.535 &1.556& 2.56 & 22.92 & 25.48 & --- & 4.45 &-&-& -298.31 & -20.49 \\
\hline
\end{tabular}
\caption{Numerical results for $k=0.389$~GeV$^3$, $m_{\rho}^{s.w.} =0.778$  GeV, $k_1=-0.78$  GeV$^3$, $k_2=0.5$~GeV$^3$, $\sigma=(0.368)^{3}$  GeV$^3$ and $ m_q=0.00145$~GeV.
\label{tab:1}}
\end{center}
\end{table}

\begin{table}[!h]
\begin{center}
\begin{tabular}{|c|c|c|c|c|c|c|c|c|c|c|c|}
\hline
n & $m_N$ (GeV) & $m_N^{s.w.}$ (GeV)&$g_{\rho NN}^{(0)}$  & $g_{\rho NN}^{(1)}$ & $g_{\rho NN}^{s.w.} $ & $g_{\rho NN}^{exp} $ & $g_{\rho NN}^{h.w.} $ & $g_{\rho NN}^{q.s.r} $  & $g_{\rho NN}^{c.q.m.} $ & $f_{\rho}^{s.w.}$ & $f_{\rho}^{h.w.}$  \\
\hline
0 & 0.94 & 1.089& 1.645 & 3.688 & 5.333 &2.52$\pm$0.06 & -8.6 & -2.5$\pm$1.1 &2.8& -16.24 & -21.10 \\
 & & & & & & 4.2$\sim$6.5 & -4.3$\sim$-6.2 &  & && \\
\hline
1 & 1.44 & 1.323 & 2.15 & 8.37 & 10.52 & --- & 25.94 &-&-& -73.85 & -20.18 \\
\hline
2 & 1.535 & 1.556 & 2.56 & 15.99 & 18.55 & --- & 4.45 &-&-& -206.61 & -20.49 \\
\hline
\end{tabular}
\caption{Numerical results for $k=0.389$~GeV$^3$, $m_{\rho}^{s.w.} =0.778$ ~GeV,  $k_1=-0.78$~GeV$^3$,    $k_2=0.5$~GeV$^3$, $\sigma=(0.326)^{3}$ ~GeV$^3$ and $ m_q=0.0023$~GeV.
\label{tab:2}}
\end{center}
\end{table}

\begin{table}[!h]
\begin{center}
\begin{tabular}{|c|c|c|c|c|c|c|c|c|c|c|c|}
\hline
n & $m_N$ (GeV) & $m_N^{s.w.}$ (GeV)  &$g_{\rho NN}^{(0)}$  & $g_{\rho NN}^{(1)}$ & $g_{\rho NN}^{s.w.} $ & $g_{\rho NN}^{exp} $ & $g_{\rho NN}^{h.w.} $ & $g_{\rho NN}^{q.s.r} $  & $g_{\rho NN}^{c.q.m.} $ & $f_{\rho}^{s.w.}$ & $f_{\rho}^{h.w.}$  \\
\hline
0 & 0.94 & 0.98 & 1.48 & 5.9 & 7.38 &2.52$\pm$0.06  & -8.6 & -2.5$\pm$1.1 &2.8& -29.81 & -21.10 \\
  & & & & & &  4.2$\sim$6.5 & -4.3$\sim$-6.2 & & && \\
\hline
1 & 1.44 &1.19 & 1.93 & 13.457 & 15.387 & --- & 25.94 &-&-& -122 & -20.18 \\
\hline
2 & 1.535 &1.14 & 2.3 & 25.654 & 27.954 & --- & 4.45 &-&-& -335 & -20.49 \\
\hline
\end{tabular}
\caption{Numerical results for $k=0.35$~GeV$^3$,  $m_{\rho}^{s.w.} =0.7$~GeV,  $k_1=-0.78$~GeV$^3$,  $k_2=0.5$~GeV$^3$, $\sigma=(0.368)^{3}$ ~GeV$^3$ and  $ m_q=0.00145$~GeV.
\label{tab:3}}
\end{center}
\end{table}

\begin{table}[!h]
\begin{center}
\begin{tabular}{|c|c|c|c|c|c|c|c|c|c|c|c|}
\hline
n & $m_N$ (GeV) & $m_N^{s.w.}$ (GeV) &$g_{\rho NN}^{(0)}$  & $g_{\rho NN}^{(1)}$ & $g_{\rho NN}^{s.w.} $ & $g_{\rho NN}^{exp} $ & $g_{\rho NN}^{h.w.} $& $g_{\rho NN}^{q.s.r} $& $g_{\rho NN}^{c.q.m} $ & $f_{\rho}^{s.w.}$ & $f_{\rho}^{h.w.}$  \\
\hline
0 & 0.94 & 0.98 & 1.48 & 4.2 & 5.68 & 2.52$\pm$ 0.06  & -8.6 & -2.5$\pm$1.1 &2.8& -19.98 & -21.10 \\
 & & & & & &  4.2$\sim$6.5 & -4.3$\sim$-6.2 & & & &  \\
\hline
1 & 1.44 &1.19 & 1.93 & 9.41 & 11.34 & --- & 25.94 &-&-& -84.33 & -20.18 \\
\hline
2 & 1.535 &1.14 & 2.3 & 17.88 & 20.18 & --- & 4.45&-&- & -202.5 & -20.49 \\
\hline
\end{tabular}
\caption{Numerical results for $k=0.35$~GeV$^3$,  $m_{\rho}^{s.w.}=0.7$~GeV,   $k_1=-0.78$~GeV$^3$,    $k_2=0.5$~GeV$^3$,   $\sigma=(0.326)^{3}$ ~GeV$^3$ and  $ m_q=0.0023$~GeV.
\label{tab:4}}
\end{center}
\end{table}

\section{summary}

In the present article we have calculated the strong coupling constant of $\rho$ mesons with nucleons within the soft-wall model of AdS/QCD. We found that the predictions of the soft-wall model for the $g_{\rho NN}$ coupling constant are closer to the experimental value than the one obtained in the hard-wall model framework. Unfortunately,  there is no experimental data for determining of the $g_{\rho NN}$ constant in the case of excited states of the nucleon and the soft-wall model results obtained here disagrees with the ones found in the hard-wall model framework in~\cite{16}. Also, the soft-wall model value for the tensorial coupling $f_{\rho}$ agrees with the hard-wall one for the ground-states of nucleons. However, the results for this coupling in the case of excited states of these particles obtained in both models are different and may be verified in future experiments.

{\bf Acknowledgements}

Authors thanks to T.M. Aliev for useful discussion of results. Sh.M. thanks to A. Mustafayev for reading manuscript. Sh.M. was supported by TUBITAK grant 2221 - Fellowships for Visiting Scientists and Scientists on Sabbatical Leave (Turkey).

\end{document}